\documentclass[11pt]{article}

\usepackage[left=2.5cm,top=4cm,right=2.5cm,bottom=3cm]{geometry}
\usepackage{amsmath,amssymb}
\usepackage{placeins}
\usepackage{float}

\usepackage{subfigure,epsfig,epsf,color}
\usepackage[utf8x]{inputenc}
\usepackage{slashed}
\usepackage{epstopdf}
\usepackage{subfigure}
\usepackage{amsfonts}
\usepackage{cite}
\usepackage{pdflscape}
\usepackage{lipsum}

\begin{document}
 \date{}

\title{Inflationary Cosmology in a non-minimal $f(R,T)$ gravity theory using a $RT$ mixing term }
\maketitle
 \begin{center}
 Payel Sarkar\footnote{p20170444@goa.bits-pilani.ac.in~(Corresponding author)},~Ashmita\footnote{p20190008@goa.bits-pilani.ac.in},~Prasanta Kumar Das\footnote{pdas@goa.bits-pilani.ac.in} \\
 \end{center}
 
 \begin{center}
 Department of Physics, \\
 Birla Institute of Technology and Science-Pilani, K. K. Birla Goa campus, \\
 NH-17B,Bypass Road, Zuarinagar, Goa-403726, India
 \end{center}
 \vspace*{0.25in}
\begin{abstract}
We investigate a class of inflationary models in modified gravity theories which contain a non-minimal coupling between gravity and a scalar field $\phi$ (inflaton) as $f(R,T)=R \bigl(1+\alpha+ \kappa^4 \beta T \bigr)+\kappa^2\gamma T $ where $\kappa^2=8\pi G$ where $G$ is the Newton's constant. We consider two inflaton potentials of the form (i) $V = V_0 \bigl(1 +\ln{\phi} \bigr)$ and (ii) $V_0\frac{\lambda \phi^p}{1+\lambda\phi^p}$. For a range of potential parameters, we explored the constraints on modified gravity parameters i.e. ($\alpha$, $\beta$ and $\gamma$) in three categories --- (i) $\beta \neq0$, $\alpha=\gamma=0$ (considering  $R$ and $RT$ mixing terms), (ii) $\alpha=0$, $\gamma\neq0$, $\beta\neq 0$ ($RT$ mixing term along with $T$ and $R$ terms) and (iii) $\gamma=0$, $\alpha\neq0$, $\beta\neq0$ ($RT$ mixing term along with $R$ term) for the above two potentials. The inclusion of $RT$ mixing term provides the scalar spectral index $n_s$ up to $3\sigma$ limit of PLANCK data, which is $n_s=0.9649\pm0.0042$ as well as the tensor-to-scalar ratio $r<0.106$ and the e-fold parameter $40<N<70$ for both the potentials.
\end{abstract}

\section{Introduction}
Although the history of the observable universe prior to the era of nucleosynthesis is unknown, it is commonly accepted that cosmic inflation occurred just after the Big Bang. 
It is a theory of exponential accelerated expansion of the universe, first envisioned by A. Guth in 1981 \cite{Einstein,Guth}, when the entire observable universe, once contained in a tiny region of the Universe all at the same temperature, suddenly expanded (in a brief fraction of a second).  \\
\noindent Inflation, which solves cosmological problems such as flatness problem, and horizon problem \cite{Kolb,Liddle}, is a hot research topic in modern cosmology. It helps us to understand ``why the observable universe today is so much isotropic and homogeneous?" Inflation also acts as the seed of density fluctuations and CMB anisotropies \cite{Kolb, Spergel}, which later evolve as Large scale structures of our universe. The scale invariant cosmological perturbations arose due to the density fluctuations, agree remarkably well with observational data such as COBE\cite{smoot}, BICEP\cite{BICEP}, PLANCK\cite{PLANCK} and WMAP\cite{WMAP}. The WMAP data measures the spectral index of the scalar fluctuations $n_s = 0.99 \pm 0.04$ and put the $95\%$ C.L. upper limit on the tensor-to-scalar ratio, $r < 0.9$. The recent PLANCK-2018 mission measures the scalar spectral index $n_s = 0.9649 \pm 0.0042$ and put the  upper limit on $r < 0.1$ (at $95\%$ C.L.), which is further tightened by combining with the BICEP2/Keck Array BK15 data to obtain $r < 0.056$.\\
Normally, an accepted model of Inflation requires a scalar field in a flat potential, which rolls slowly over the potential for  a sufficiently long period of time\cite{Liddle,Linde1,Linde2, Baumann, Kinney, Gron,sarkar1}. But the tightly constrained tensor-to-scalar ratio $r$ rules out many such inflationary models. Besides, a viable inflationary model has to produce the necessary e-fold($N$) to trigger inflation and predict a smaller tensor-scalar ratio $r$ in order to fit with the observational data.\\
The simplest extension of the scalar field Lagrangian is to introduce a coupling between scalar matter and gravity. The non-minimal inflation model describes the cosmic expansion with a graceful exit \cite{Nozari} towards its end \cite{cheong, Jin}. It is also intriguing to consider a non-minimal coupling scenario in the context of multidimensional theories like super-string theory \cite{Maeda} and induced gravity \cite{Accetta2}. In presence of non-minimal matter-gravity term, the conservation equation may not be valid anymore and the extra term in energy-momentum leads to an irreversible particle creation process, which appears in quantum field theory in curved space-time\cite{Faraoni, Faraoni2,Dzhunushaliev,Dzhunushaliev1, Lobato, Chen, Parker1, Parker2}. The addition of non-minimal matter-gravity coupling in modified gravity theory is widely investigated nowadays to explain the late-time acceleration by making use of matter-gravity coupling without introducing any dark matter component.\\
A host of modified gravity have been studied such as $f(R,T)$ gravity \cite{Harko,Roshan}, $f(R)$gravity\cite{Bertolami,Bisabr, Sahoo}, $f(R,\mathcal{L}_m)$ gravity \cite{Nojiri1, Allemandi}, $f(\phi,T)$\cite{chen2, Nisha} along this direction. 
In this manuscript, we have mentioned two different types of inflationary models in modified $f(R,T)$ gravity where $T$ (the trace of energy-momentum tensor) is coupled with the Ricci scalar $R$. We have taken a modified gravity model with $f(R,T)=R(1+\alpha+\kappa^4\beta T)+\kappa^2\gamma T$\cite{Chen3} to study inflationary expansion. Here $\alpha, \beta$ and $\gamma$ are the modified gravity parameters which will be estimated using WMAP and PLANCK 2018 data for both the models using a class of inflaton potentials.   
Chaotic and natural inflation models have also been studied using the aforementioned functional form of $f(R,T)$\cite{Chen3}, but no such work has been done for fractional and logarithmic inflaton potential. \\
\noindent In this work, we have considered logarithmic and fractional potential having the form  (i) $V=V_0(1+\ln{\phi})$ and (ii) $V=V_0\frac{\lambda\phi^p}{1+\lambda\phi^p}$. The analysis of the slow-roll inflation for $V=V_0\frac{\lambda\phi^p}{1+\lambda\phi^p}$ has been reported by these authors in their earlier works \cite{sarkar1,Ashmita, Ashmita2} for $f(R,T)$ gravity with both linear and higher order term of $T$ and for non-minimal coupling term $\frac{1}{2}\xi R\phi^2$ whereas the potential $V=V_0(1+\ln{\phi})$ have been discussed in \cite{Gron} to study inflation. Hence to, study the inflationary expansion in modified gravity theory with the $RT$ mixing term for the above two potentials, is a worthwhile exercise. \\ 
\noindent
The paper is organized as follows. In section 2, we obtain the Einstein Field equations in the modified $f(R,T)$ gravity and derive the potential slow-roll parameters in this $f(R,T)$ gravity. In section 3, we discuss the inflationary scenario for two different potentials and derive the cosmological parameters such as scalar spectral index $n_s$ and tensor-to-scalar ratio $r$. These cosmological parameters have been subject to constraints in the parameter space of inflaton potential $(\lambda,p)$ within the context of modified $f(R,T)$ gravity. We analyze our results and compare those  with the PLANCK 2018\cite{PLANCK} and the WMAP\cite{WMAP} data. In section 4, we briefly summarize our results and conclude.
\section{Field equations in modified gravity with RT mixing term}
In this modified gravity model where the inflaton field couples non-minimally with gravity, the action is given by,
\begin{equation}\label{action}
    \mathcal{S}=\int d^4x\sqrt{-g}\left[\frac{f(R,T)}{2\kappa^2}+\mathcal{L}_m\right]
\end{equation}
where $\kappa^2=8\pi G=\frac{1}{M_{Pl}^2}$ \footnote{In our analysis, we have set $\kappa^2=8\pi G=1$}, $g$ is the determinant of $g_{\mu\nu}$, $R$ is Ricci scalar and $T$ is the trace of energy-momentum tensor $T_{\mu\nu}$. The form of $f(R,T)$ considered here as 
$$f(R,T)=R \bigl(1+\alpha+\kappa^4\beta T \bigr)+\kappa^2\gamma T$$ where $\alpha,~\beta$ and $\gamma$ are the modified gravity parameters. One can recover the Einstein gravity by taking the limit $\alpha,\beta,\gamma$ $\rightarrow 0$. The energy-momentum tensor of the scalar field can be written as,
\begin{equation}
T_{\mu\nu}=\partial_{\mu}\phi\partial_{\nu}\phi+g_{\mu\nu}\mathcal{L}_m
\end{equation}
where $\mathcal{L}_m=-\frac{1}{2}g^{\mu\nu}\partial_{\mu}\phi\partial_{\nu}\phi-V(\phi)$ is the Lagrangian of the scalar field, $V(\phi)$ is the scalar potential.
\noindent By varying the action (i.e. Eq. (\ref{action})) with respect to metric $g_{\mu\nu}$, the modified Einstein field equations can be obtained as
\begin{equation}\label{field}
    f_R R_{\mu\nu}-\frac{1}{2}f g_{\mu\nu}+\bigl(g_{\mu\nu}\square-\nabla_{\mu}\nabla_{\nu} \bigr)f_R=T_{\mu\nu}-f_T \bigl(T_{\mu\nu}+\Theta_{\mu\nu} \bigr)
\end{equation}
Here $f_R = \partial f/\partial R$, $f_T = \partial f/\partial T$ and $\Box = \nabla_\mu \nabla^\mu$, where $\nabla_\mu$ is the covariant derivative. The tensor $\Theta_{\mu\nu}$ is defined in terms of the energy-momentum tensor as,
\begin{equation}
    \Theta_{\mu\nu}= g^{\lambda \kappa} \frac{\delta T_{\lambda \kappa}}{\delta g^{\mu\nu}}=-T_{\mu\nu}-\partial_{\mu}\phi\partial_{\nu}\phi
\end{equation}
Using the above form of $f(R,T)$, the modified Einstein equation can be written as,
\begin{equation}\label{Einstein}
    R_{\mu\nu}-\frac{1}{2}Rg_{\mu\nu}=T_{\mu\nu}^{eff}
\end{equation}
 where $T_{\mu\nu}^{eff}$, the effective energy-momentum tensor, is given by
 \begin{equation}
 \label{Tmunu}
    T_{\mu\nu}^{eff}=T_{\mu\nu}+\bigl(R \beta+\gamma \bigr)\partial_{\mu}\phi\partial_{\nu}\phi-R_{\mu\nu} \bigl(\alpha+\beta T \bigr)+\frac{1}{2}g_{\mu\nu} \Bigl\{R \bigl(\alpha+\beta T \bigr) + \gamma T\Bigr\}-\bigl(g_{\mu\nu}\square-\nabla_{\mu}\nabla_{\nu} \bigr)\beta T
 \end{equation}
Accordingly, the effective energy density and the effective pressure can be derived from the energy-momentum tensor as,
\begin{equation}
    \rho^{eff}=\frac{1}{2}\dot{\phi}^2 \Bigl(1+12\beta\dot{H}+\gamma+18\beta H^2 \Bigr)+V \Bigl(1+2\gamma+12\beta H^2 \Bigr)-3\alpha H^2-6H\beta\dot{\phi}\ddot{\phi}+12\beta H\dot{\phi}V_{,\phi}
\end{equation}
and
\begin{equation}
\begin{split}
    p^{eff}=&\frac{1}{2}\dot{\phi}^2 \Bigl(1+\gamma+4\beta\dot{H}+6\beta H^2 \Bigr)-V \Bigl(1+2\gamma+8\beta\dot{H}+12\beta H^2 \Bigr)+2\alpha\dot{H}+3\alpha H^2+\beta \Bigl(2\ddot{\phi}^2+2\dot{\phi}\dddot{\phi}\\
    & -4\ddot{\phi}V_{,\phi}-4V_{,\phi\phi}\dot{\phi}^2+4H\dot{\phi}\ddot{\phi}-8H\dot{\phi}V_{,\phi} \Bigr)
    \end{split}
\end{equation}
Solving the Einstein field equation, Eq. (\ref{Einstein}) along with Eq. (\ref{Tmunu}), we obtain the modified Friedmann equation and acceleration equation as
\begin{equation}
\label{friedmann1}
    3H^2 \bigl(1+\alpha \bigr)=\frac{1}{2}\dot{\phi}^2 \bigl(1+12\beta\dot{H}+\gamma+18\beta H^2 \bigr)+V \bigl(1+2\gamma+12\beta H^2 \bigr)
\end{equation}
\begin{equation}
\label{friedmann2}
\begin{split}
    \frac{\ddot{a}}{a} \bigl(1+\alpha \bigr)=
    &-\frac{1}{3}\biggl\{\dot{\phi}^2\Bigl(1+\gamma+6\beta\dot{H}-6\beta V_{,\phi\phi}+9\beta H^2 \Bigr)-V\Bigl(1+2\gamma+12\beta\dot{H}+12\beta H^2\Bigr) \biggr\}\\
    & \qquad -\beta H\dot{\phi}\ddot{\phi}+2\beta H\dot{\phi}V_{,\phi}-\beta\ddot{\phi}^2-\beta\dot{\phi}\dddot{\phi}+2\beta\ddot{\phi}V_{,\phi}
    \end{split}
\end{equation}
Furthermore, the equation of continuity of the inflaton takes the form:
\begin{equation}
\label{EOM}
    \ddot{\phi}\Bigl(1+\gamma+6\beta\dot{H}+12\beta H^2\Bigr)+V_{,\phi}\Bigl(1+2\gamma+12\beta\dot{H}+24\beta H^2\Bigr)+3H\dot{\phi}\Bigl(1+\gamma+10\beta H^2+8\beta \dot{H}\Bigr)+6\beta\dot{\phi}\frac{\dddot{a}}{a}=0
\end{equation}
In the limit $\alpha \to 0$, $\beta \to 0$ and $\gamma \to 0$, the above set of modified equations take the form 
$$ 3H^2 = \frac{1}{2}\dot{\phi}^2+V(\phi),~~ \frac{\ddot{a}}{a} = -\frac{1}{3}\Biggl[\dot{\phi}^2 - V \Biggr]  \rm{and} ~~ \ddot{\phi}+3H\dot{\phi}+V_{,\phi}=0$$
These are the Friedmann, acceleration and continuity equations in Einstein gravity with a scalar field.
\subsection{Inflation: Slow-roll conditions} 
The slow-roll conditions required for inflationary expansion, are listed as follows
$$\dot{\phi}^2<<V(\phi), |\dddot{\phi}|<<|H\ddot{\phi}|<<|H^2\dot{\phi}|,~~|\ddot{H}|<<|H\dot{H}|<<|H^3|$$ 
Applying these conditions, we can approximate Eq. (\ref{friedmann1}) and Eq. (\ref{friedmann2}) as,
\begin{equation}
\label{Hubble}
   H^2  \approx \frac{V}{3}\Biggl[\frac{1+2\gamma}{1+\alpha-4\beta V}\Biggr]
\end{equation}
 and
 \begin{equation}
 \label{acceleration}
 \dot{H}\approx -\frac{1}{2}\dot{\phi}^2\Biggl[\frac{(1+\alpha)(1+\gamma)+4\beta V\gamma}{(1+\alpha-4\beta V)^2}\Biggr]
 \end{equation}
and the equation of continuity (Eq. \ref{EOM}) as 
 \begin{equation}
 \label{continuity}
     3H\dot{\phi}\Bigl\{(1+\gamma)(1+\alpha)+4\beta V\gamma\Bigr\}+V_{,\phi}(1+2\gamma)(1+\alpha+4\beta V)\approx 0
 \end{equation}
Next, we are to write all the above three equations in the Einstein frame. We use three conformal transformations to do that, which are $\Omega_1(\phi),~\Omega_2(\phi)$ and $\Omega_3(\phi)$, respectively on the metric $g_{\mu\nu}$, the inflaton field $\phi$ and the inflaton potential $V(\phi)$.\\
\noindent The first conformal transformation $\Omega_1(\phi)$ stands for transformation of metric $g_{\mu\nu}$ to an auxiliary metric $\tilde{g}_{\mu\nu}$ as,
\begin{equation}
    \tilde{g}_{\mu\nu}=\Omega_1(\phi)g_{\mu\nu}
\end{equation}
The line element can be represented as,
\begin{equation}
    \tilde{g}_{\mu\nu}dx^{\mu}dx^{\nu}=-d\tilde{t}^2+\tilde{a}(\tilde{t})^2d\vec{x}^2=\Omega_1(\phi)\left(-dt^2+a^2(t)d\vec{x}^2\right)
\end{equation}
The second scalar field transformation $\Omega_2(\phi)$, which transforms $\phi$ to an auxiliary scalar field $\tilde{\phi}$, is defined as,
\begin{equation}
    \Biggl[\frac{d\tilde{\phi}}{d\phi}\Biggr]^2=\Omega_2(\phi)
\end{equation}
And finally, the third transformation $\Omega_3(\phi)$ is used for the transformation of the scalar potential and it is defined as
\begin{equation}
    \tilde{V}(\tilde{\phi})=\Omega_3(\phi)V(\phi)
\end{equation}
Applying theses transformations on the metric, the scalar field, and the scalar potential, we can write Eqs.(\ref{Hubble}), (\ref{acceleration}) and (\ref{continuity}) in the Einstein frame as,
\begin{equation}
\tilde{H}^2\Omega_1(1+\alpha-4\beta V)=\frac{V}{3}(1+2\gamma)\longrightarrow
    3\tilde{H}^2\approx \tilde{V}
\end{equation}
\begin{equation}
\dot{\tilde{H}}\Omega_1(1+\alpha-4\beta V)=-\frac{1}{2}\dot{\phi}^2 \Biggl[\frac{(1+\alpha)(1+\gamma)+4\beta V\gamma}{1+\alpha-4\beta V}\Biggr]\longrightarrow
    \frac{d\tilde{H}}{d\tilde{t}}\approx -\frac{1}{2}\Biggl[\frac{d\tilde{\phi}}{d\tilde{t}}\Biggr]^2
\end{equation}
\begin{equation}
    3\tilde{H}\frac{d\tilde{\phi}}{d\tilde{t}}+\tilde{V}_{,\tilde{\phi}}\approx 0
\end{equation}
where $\tilde{V}_{,\tilde{\phi}}=\frac{d\tilde{V}}{d\tilde{\phi}}$ and $\tilde{H}=\frac{\dot{\tilde{a}}}{\tilde{a}}=H\Omega_1^{-1/2}$, $\dot{H}=\Omega_1\dot{\tilde{H}},~\dot{\phi}=\frac{d\tilde{\phi}}{d\tilde{t}}\Omega_1^{1/2}\Omega_2^{-1/2}$. The above set of equations can be obtained by choosing 
\begin{equation}
    \Omega_1=1+\alpha-4\beta V
\end{equation}
\begin{equation}
    \Omega_2=\frac{(1+\alpha)(1+\gamma)+4\beta\gamma V}{(1+\alpha-4\beta V)^2}
\end{equation}
and,
\begin{equation}
    \Omega_3=\frac{1+2\gamma}{(1+\alpha-4\beta V)^2}
\end{equation}
In Einstein frame, the potential slow roll parameters can be defined as,

\begin{eqnarray}\label{slow roll}
\tilde{\epsilon}_v = \frac{1}{2}\Biggl[\frac{\tilde{V}'(\tilde{\phi})}{\tilde{V}(\tilde{\phi})}\Biggr]^2,  
~~\tilde{\eta}_v = \frac{\tilde{V}^{''}(\tilde{\phi})}{\tilde{V}(\tilde{\phi})}
\end{eqnarray}
where \begin{equation*}
    \tilde{V}'(\tilde{\phi})=\frac{d\tilde{V}}{d\tilde{\phi}}=V_{,\phi}\Omega_3\Omega_2^{-1/2}, ~~
    \tilde{V}''(\tilde{\phi})=\frac{d^2\tilde{V}}{d\tilde{\phi}^2}=V_{,\phi\phi}\frac{\Omega_3}{\Omega_2}+V_{,\phi}\frac{1}{\Omega_2}\frac{d\Omega_3}{d\phi}-V_{,\phi}\frac{\Omega_3}{2\Omega_2^2}\frac{d\Omega_2}{d\phi}
    \end{equation*}
    The slow-roll parameters are related to the CMBR observables - the scalar spectral index($n_s$), tensor spectral index($n_T$) and tensor-to scalar ratio($r$) as follows,

\begin{equation}\label{spectral index}
    n_s-1=2\tilde{\eta}_v-6\tilde{\epsilon}_v, ~~ r=16 \tilde{\epsilon}_v
\end{equation}
The e-fold number $N$ is defined as the ratio of the final value of the scale factor $a_f$ during the inflationary era and its initial value $a_i$ can be calculated as, 
\begin{equation}
    N= \int_{\tilde{a_i}}^{\tilde{a_f}}\frac{d\tilde{a}}{\tilde{a}} = \int_{\tilde{\phi_i}}^{\tilde{\phi_f}} \frac{\tilde{H}}{d\tilde{\phi}/d\tilde{t}}d\tilde{\phi}
    =\int_{\phi_i}^{\phi_f} \frac{H}{\dot{\phi}}d\phi
\end{equation}

\section{Different Inflationary models}
\subsection{Model I: $V = V_0 \bigl(1 +\ln{\phi} \bigr)$: Results and Analysis} 
First, we consider the inflaton potential of the form,
\begin{equation}
    V = V_0 \bigl(1 +\ln{\phi} \bigr)
\end{equation}
where $\phi$ is in unit of $M_{Pl}$, $V_0$ is a constant expressed in units of $M_{Pl}^4$.
This potential was studied by Lyth and Riotto in \cite{Lyth} and is called ``Spontaneous broken SUSY Inflation". Apart from this, the aforementioned potential is also examined by Guo and Zhang \cite{Guo}, Hebecker {\it et al.}\cite{Hebecker}.
The slow-roll parameters for the given potential is
\begin{equation}
    \epsilon_{\Tilde{V}} = \frac{1}{ 2 \Bigl\{ \bigl(1+\alpha \bigr) \bigl(1+\gamma \bigr)+4 \beta V_0 \bigl(1 +\ln{\phi} \bigr)\Bigr\}} \times 
    \Biggl[\frac{ 1+ \alpha -4 \beta V_0 (1 +\ln{\phi})}{ \phi \bigl(1 +\ln{\phi} \bigr)} \Biggr]^2
\end{equation}

\begin{equation}
\begin{split}
      \eta_{\Tilde{V}} 
      &= -\frac{\Bigr\{1+\alpha -4 \beta V_0 \bigl(1 +\ln{\phi} \bigr)\Bigr\}}{\phi ^2 \bigl(1 +\ln{\phi} \bigr) \Bigl\{ \bigl(1+\alpha \bigr) \bigl(1+\gamma \bigr)+4 \beta V_0 \bigl(1 +\ln{\phi} \bigr)\Bigr\}^2} \times  \Biggl[\bigl(\alpha +1 \bigr)^2 \bigl(\gamma +1 \bigr)-2\beta V_0 \bigl(\alpha +1 \bigr) \bigl(2 \gamma \ln{\phi}\\
      & \qquad +4 \gamma +1 \bigr)-8 \beta ^2 V_0^2 \bigl(2 {\ln{\phi}}^2 +7 \ln{\phi}+5\bigr)\Biggr]
\end{split}
\end{equation}
The spectral index parameters are defined as follows:
\begin{equation}
\begin{split}
    n_s 
    &= 1 - \frac{3\Bigl\{ 1+ \alpha -4 \beta V_0 (1 +\ln{\phi})\Bigr\}^2}{\phi ^2 \bigl(1 +\ln{\phi} \bigr)^2 \Bigl\{ \bigl(1+\alpha \bigr) \bigl(1+\gamma \bigr)+4 \beta V_0 \bigl(1 +\ln{\phi} \bigr)\Bigr\}} - \frac{2\Bigr\{1+\alpha -4 \beta V_0 \bigl(1 +\ln{\phi} \bigr)\Bigr\}}{\phi ^2 \bigl(1 +\ln{\phi} \bigr) \Bigl\{ \bigl(1+\alpha \bigr) \bigl(1+\gamma \bigr)+4 \beta V_0 \bigl(1 +\ln{\phi} \bigr)\Bigr\}^2} \\ 
    & \qquad \times \Biggl[\bigl(\alpha +1 \bigr)^2 \bigl(1+\gamma \bigr)-2\beta V_0 \bigl(1+\alpha \bigr) \bigl(1+ 4\gamma + 2 \gamma \ln{\phi} \bigr)-8 \beta ^2 V_0^2 \bigl(5+ 7 \ln{\phi} +2 {\ln{\phi}}^2 \bigr)\Biggr]
\end{split}
\end{equation}
\begin{equation}
    r = \frac{8}{\Bigl\{ \bigl(1+\alpha \bigr) \bigl(1+\gamma \bigr)+4 \beta V_0 \bigl(1 +\ln{\phi} \bigr)\Bigr\}} \times 
    \Biggl[\frac{ 1+ \alpha -4 \beta V_0 (1 +\ln{\phi})}{ \phi \bigl(1 +\ln{\phi} \bigr)} \Biggr]^2
\end{equation}
\noindent In Table~(\ref{table:1a}), we have shown the range of modified gravity parameters along with the values of different cosmological parameters for $V=V_0\left(1+\ln{\phi}\right)$. 
\begin{table}[htb]
\centering
\small
\addtolength{\tabcolsep}{0.5pt}
\begin{tabular}{cccccccc}
\hline
For potential $V = V_0 \bigl(1 +\ln{\phi} \bigr)$, & $\alpha = 0$, & $\gamma = 0$ & & & & & \\
Range of $\beta$ & $\beta$ &  $\phi $ &  $\phi_f$ & N & $n_s$ & r &\\ 
$-0.018690 < \beta < 0.003127$ & -0.001711 & 6.5 & 0.850730 & 49 & 0.973331 & 0.024349 &\\

\hline
When $\alpha = 0$, & $\gamma \ne 0$ & & & & & & \\
Range of $\gamma$ & $\gamma $ & $\beta$ & $\phi $ &  $\phi_f$ & N & $n_s$ & r \\  
$-0.4802 < \gamma < -0.158$ & -0.3338 & -0.01 & 8 & 0.963083 & 47 & 0.96497 & 0.0517803\\

$-0.1573 < \gamma < 0.3151$ & 0.1087 & 0.001 & 6 & 0.826166 & 46 & 0.973328 & 0.024895\\
\hline

When $\gamma = 0$, & $\alpha \ne 0$ & & & & & & \\
Range of $\alpha$ & $\alpha $ & $\beta$ & $\phi $ &  $\phi_f$ & N & $n_s$ & r \\  
$-0.3068 < \alpha < 0.4654$ & 0.0862 & -0.01 & 6.7 & 0.883085 & 45 & 0.965013 & 0.031529\\

$-0.2315 < \alpha < 0.1819 $ & -0.09369 & 0.001 & 6 & 0.82681 & 46 & 0.973322 & 0.024912\\
\hline
\end{tabular}
\caption{\label{table:1a} For $V = V_0 \bigl(1 +\ln{\phi} \bigr)$, the e-fold number $N$ and the spectral index parameters $n_s$ and $r$ calculated for a fixed value of $\phi$ and $\alpha$ are presented. Here $\phi$ is rescaled as $\frac{\phi}{M_{Pl}}$}
\end{table}
\noindent
\noindent With $RT$ mixing term only, for $\beta = -0.001711$, we find the e-fold parameter $N=49$ and the  spectral index $n_s = 0.973331$ and $r=0.024349$, while by taking $\alpha=0$ and $\gamma = -0.3338(0.1087)$ and  $\beta = -0.010(0.001)$, we find $N=47(46)$ and $n_s = 0.96497(0.97333)$ and $r=0.05178(0.024805)$. Lastly, for $\gamma = 0$, with $\alpha = 0.0862(-0.09369)$ and $\beta = -0.01(0.001)$, we find $N=45(46)$ and $n_s = 0.965013(0.973322)$ and $r=0.031529(0.024912)$.
We see that the values of the scalar spectral index $n_s$ and the tensor-to-scalar ratio $r$ for the particular value of $\beta$, $\gamma$, and $\alpha$ chosen from the range e.g. $-0.018690 \le \beta \le 0.003127$, $-0.4802(-0.1573) \le \gamma \le -0.158(0.3151$ and $-0.3068(-0.2315) \le \alpha \le 0.4654(0.1819)$ brings it to a better agreement with observational PLANCK 2018 data and WMAP data along with e-fold number $N$ lying between $40-70$.
\begin{figure}[htb]
 \centerline {\psfig{file=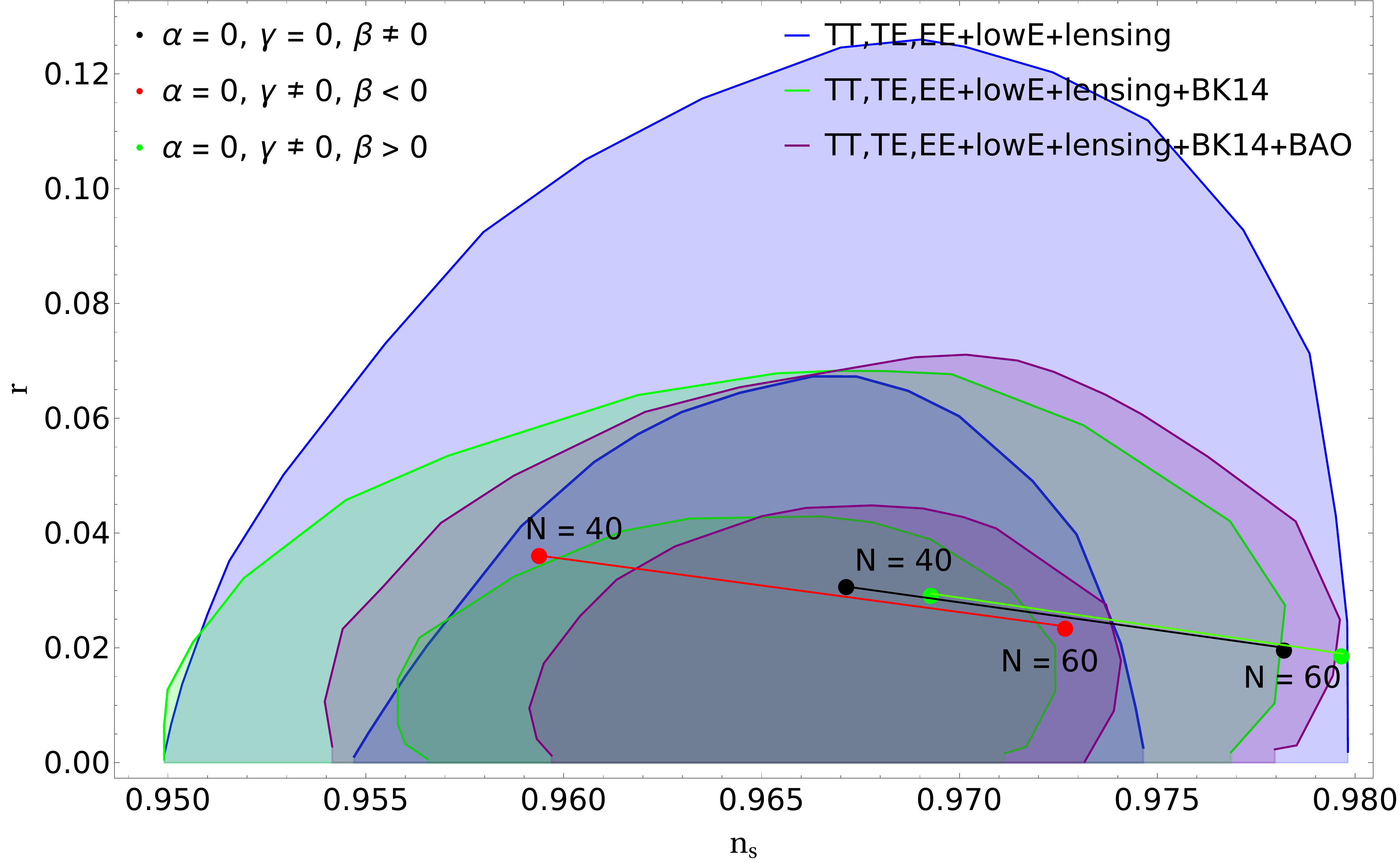,width=8.5cm}\psfig{file=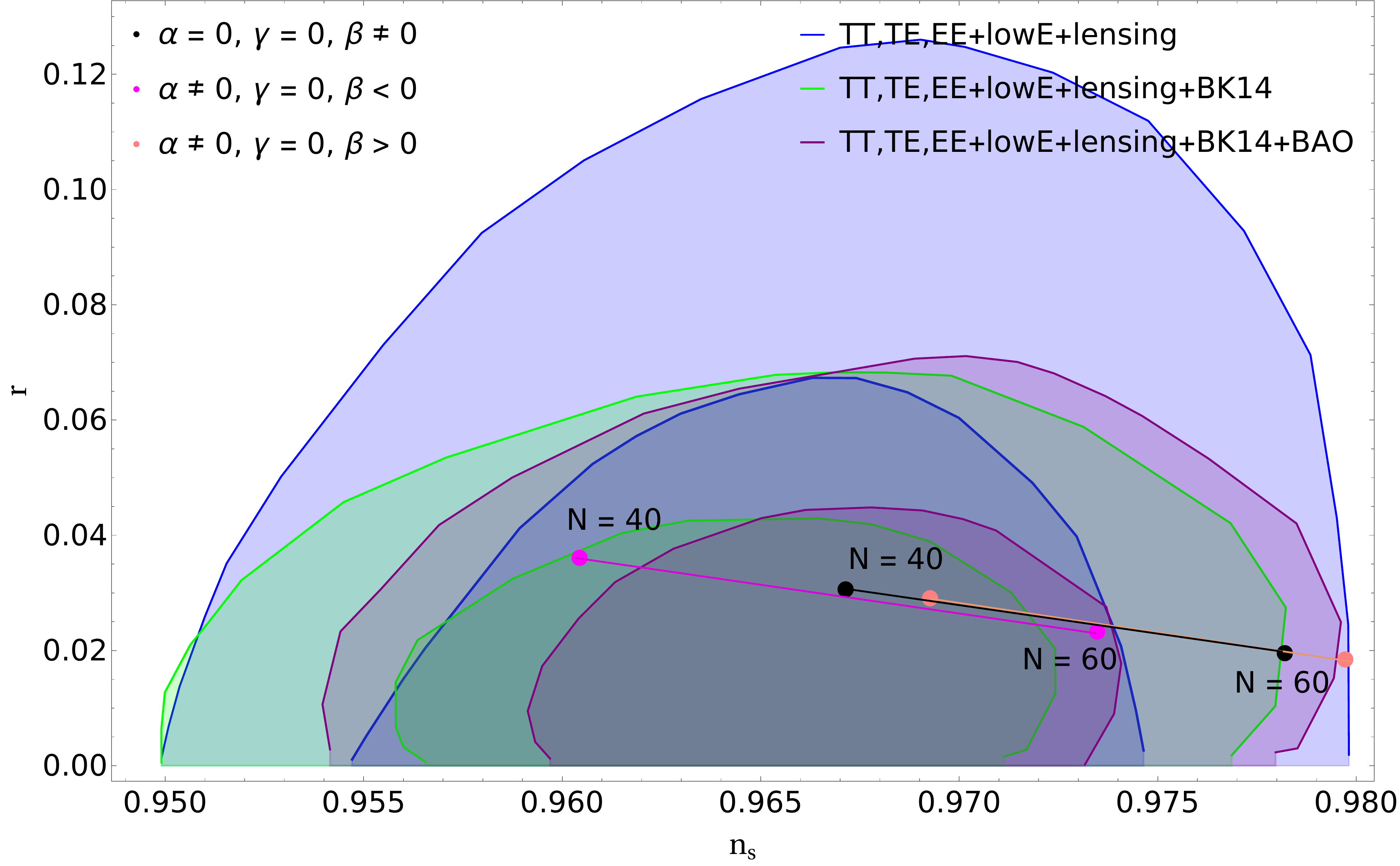,width=8.5cm}}
 \vspace*{3pt}
  \caption{(Color online) Constraints on $n_s$ and $r$ from CMB measurements of different potential. PLANCK alone, PLANCK+BK15, PLANCK+BK15+BAO upto $68\%$ Confidence Level. \label{Plot1}}
\end{figure}

\noindent
In Figure \ref{Plot1}, we have plotted the contour plot in the $n_s$--r plane for different values of $\alpha$, $\beta$, and $\gamma$, respectively. To see the effect of the non-minimally coupled term i.e. the $RT$ mixing term, first, we switch off the other two terms (by fixing $\alpha=0$ and $\gamma =0$). The blue, green, and purple regions show the contours of recent Planck 2018 in  $n_S-r$ plane. The red line indicates the plot corresponding to the non-zero $\beta$ value for $N=40$ and $60$. We have shown the effect of a fixed value of positive and negative $\beta$ for two cases (i) $\alpha = 0$ and $\gamma \ne 0$ (ii) $\alpha \ne 0$ and $\gamma = 0$. The left-hand side of Figure \ref{Plot1} shows the effect for $\beta<0$ while the right-hand side of the same figure represents the effect for the $\beta>0$ case. The black line in the plot represents the case where $\alpha \ne 0$, $\gamma = 0$ and $\beta<0$ while the blue line represents the case where $\alpha = 0$, $\gamma \ne 0$. All three cases match suitably well with the Planck 2018 data as far as the left-hand side of the plot. The right-hand side of the plot is shown for the positive $\beta$ value. We also see the same two cases (i) $\alpha = 0$ and $\gamma \ne 0$ (ii) $\alpha \ne 0$ and $\gamma = 0$ for $\beta>0$. Here the green line indicates the case when $\alpha \ne 0$ and $\gamma = 0$ while the purple line indicates $\alpha =0$ and $\gamma \ne 0$ case. We can see that as $\beta>0$, the scalar spectral index value shifts towards the $+3\sigma$ value of $n_s$ while for the $\beta<0$ case, the $n_s$ value is within $\pm 3\sigma$ range. Both cases match well with the Planck 2018 $n_s - r$ predictions.
\subsection{Model II: $V = V_0 \bigl(\frac{\lambda \phi^p}{1+ \lambda \phi^p}\bigr)$: Results and Analysis }
Next, we consider the fractional potential for inflationary expansion,
\begin{equation}
    V=V_0\frac{\lambda\phi^p}{1+\lambda\phi^p}
\end{equation}
where $V_0$ is a constant, $p$ and $\lambda$ are the potential parameters. In our analysis, we have taken the potential parameters $p = 2$, and $\lambda=1,2$, respectively. With $p=2$, the fractional potential was first studied by Eshagli et al.\cite{Eshagli}. This potential have also been studied in the minimal scenario in normal Einstein gravity and non-minimal coupled gravity\cite{sarkar1}. \\
The modified potential slow-roll parameters for this potential are found to be 
\begin{equation}
    \epsilon_{\Tilde{V}} = \frac{1}{2 \Bigl\{ \bigl(1+\alpha \bigr)\bigl(1+\gamma \bigr) +\frac{4 \beta  \lambda V_0 \phi ^p}{1+\lambda \phi^p} \Bigr\}} \times \Biggl[\frac{p \bigl(1+\alpha -\frac{4 \beta  \lambda V_0 \phi ^p}{1+\lambda \phi^p}\bigr)}{\phi \bigl(1+\lambda \phi^p \bigr) } \Biggr]^2
\end{equation}

\begin{equation}
\begin{split}
    \eta_{\Tilde{V}} 
    &= -\frac{1}{\phi^2 (1+\lambda \phi^p)^3 \Bigl\{ 1+\gamma +\lambda \phi^p +4 V_0 \beta \lambda \phi^p +\gamma \lambda \phi^p +\alpha \bigl(1+\gamma \bigr) \bigl(1+ \lambda \phi^p \bigr) \Bigr\}^2} \Biggl[ p~\Bigl\{ 1+\alpha +\alpha \lambda \phi^p - \\
    & \qquad \bigl(-1 + 4V_0 \beta \bigr) \lambda \phi^p \Bigr\} \times \biggl[ \bigl( 1+ \lambda \phi^p \bigr) \Bigl\{ 1+\alpha +\alpha \lambda \phi^p - \bigl(-1 + 4V_0 \beta \bigr) \lambda \phi^p \Bigr\} \Bigl\{ 1+\gamma +\lambda \phi^p +4 V_0 \beta \lambda \phi^p \\
    & \qquad  +\gamma \lambda \phi^p +\alpha \bigl(1+\gamma \bigr) \bigl(1+ \lambda \phi^p \bigr) \Bigr\} +p \biggl\{ -1-\lambda \phi^p -2 V_0 \beta \lambda \phi^p +\lambda^2 \phi^{2p} -2 V_0 \beta \lambda^2 \phi^{2p}-8 V_0^2 \beta^2 \lambda^2 \phi^{2p} \\
    & \qquad + \lambda^3 \phi^{3p} -16 V_0^2 \beta^2 \lambda^3 \phi^{3p} + \bigl( 1+\gamma \bigr) \bigl(-1+\lambda \phi^p \bigr) \bigl( \alpha +\alpha \lambda \phi^p \bigr)^2 -\gamma \bigl( 1+\lambda \phi^p\bigr) \Bigl\{ 1+ \bigl(-1 +4 V_0 \beta \bigr) \lambda^2 \\
    & \qquad \times \phi^{2p} \Bigr\} -2 \alpha \bigl( 1+\lambda \phi^p \bigr) \Bigl\{ 1+V_0\beta \lambda \phi^p -\lambda^2 \phi^{2p} +\gamma \bigl\{ 1+ (-1+2 V_0 \beta) \lambda^2 \phi^{2p} \bigr\} \Bigr\} \biggr\} \biggr] \Biggr] \\
    \end{split}
\end{equation}
The scalar spectral index and tensor to scalar ratio can be obtained as,
\begin{equation}
\begin{split}
    n_s 
    &= 1 - \frac{3}{\Bigl\{ \bigl(1+\alpha \bigr)\bigl(1+\gamma \bigr) +\frac{4 \beta  \lambda V_0 \phi ^p}{1+\lambda \phi^p} \Bigr\}} \times \Biggl[\frac{p \bigl(1+\alpha -\frac{4 \beta  \lambda V_0 \phi ^p}{1+\lambda \phi^p}\bigr)}{\phi \bigl(1+\lambda \phi^p \bigr) } \Biggr]^2 -\frac{2}{\phi^2 (1+\lambda \phi^p)^3} \times \\ 
    & \qquad \frac{1}{ \Bigl\{ 1+\gamma +\lambda \phi^p +4 V_0 \beta \lambda \phi^p +\gamma \lambda \phi^p +\alpha \bigl(1+\gamma \bigr) \bigl(1+ \lambda \phi^p \bigr) \Bigr\}^2} \Biggl[ p~\Bigl\{ 1+\alpha +\alpha \lambda \phi^p - \bigl(-1 + 4V_0 \beta \bigr) \lambda \phi^p \Bigr\} \\
    & \qquad  \times \biggl[ \bigl( 1+ \lambda \phi^p \bigr) \Bigl\{ 1+\alpha +\alpha \lambda \phi^p - \bigl(-1 + 4V_0 \beta \bigr) \lambda \phi^p \Bigr\} \Bigl\{ 1+\gamma +\lambda \phi^p +4 V_0 \beta \lambda \phi^p +\gamma \lambda \phi^p +\alpha \bigl(1+\gamma \bigr) \times \\
    & \qquad   \bigl(1+ \lambda \phi^p \bigr) \Bigr\} +p \biggl\{ -1-\lambda \phi^p -2 V_0 \beta \lambda \phi^p +\lambda^2 \phi^{2p} -2 V_0 \beta \lambda^2 \phi^{2p}-8 V_0^2 \beta^2 \lambda^2 \phi^{2p} + \lambda^3 \phi^{3p} -16 V_0^2 \beta^2 \lambda^3  \\
    & \qquad \times \phi^{3p}+ \bigl( 1+\gamma \bigr) \bigl(-1+\lambda \phi^p \bigr) \bigl( \alpha +\alpha \lambda \phi^p \bigr)^2 -\gamma \bigl( 1+\lambda \phi^p\bigr) \Bigl\{ 1+ \bigl(-1 +4 V_0 \beta \bigr) \lambda^2 \phi^{2p} \Bigr\} -2 \alpha \bigl( 1+\lambda \phi^p \bigr) \\
    & \qquad \times \Bigl\{ 1+V_0\beta \lambda \phi^p -\lambda^2 \phi^{2p} +\gamma \bigl\{ 1+ (-1+2 V_0 \beta) \lambda^2 \phi^{2p} \bigr\} \Bigr\} \biggr\} \biggr] \Biggr] \\
    \end{split}
\end{equation}

\begin{equation}
    r = \frac{8}{\Bigl\{ \bigl(1+\alpha \bigr)\bigl(1+\gamma \bigr) +\frac{4 \beta  \lambda V_0 \phi ^p}{1+\lambda \phi^p} \Bigr\}} \times \Biggl[\frac{p \bigl(1+\alpha -\frac{4 \beta  \lambda V_0 \phi ^p}{1+\lambda \phi^p}\bigr)}{\phi \bigl(1+\lambda \phi^p \bigr) } \Biggr]^2
\end{equation}
\begin{table}[htb]
\centering
\small
\addtolength{\tabcolsep}{0.5pt}
\begin{tabular}{cccccccccc}
\hline
For potential $V = V_0 \bigl(\frac{\lambda \phi^p}{1+ \lambda \phi^p}\bigr)$, & $\alpha = 0$, & $\gamma = 0$, & $p=2$ & &  & & & \\
Range of $\beta$ & $\lambda$  & $\beta$ &  $\phi $ &  $\phi_f$ & N & $n_s$ & r &\\ 
$-0.04 < \beta < 0.0171$ & 1  & -0.01 & 4.5 & 0.845504 & 54 & 0.969229 & 0.00392 &\\

$-0.07 < \beta < -0.008$ & 2  & -0.04 & 4.0 & 0.752338 & 59 & 0.964436 & 0.00290 &\\ 
\hline
When $\alpha = 0$, & $\gamma \ne 0$,  & $p=2$& & & & \\

Range of $\gamma$ &$\lambda$  &  $\gamma $ & $\beta$ & $\phi $ &  $\phi_f$ & N & $n_s$ & r \\  

$-0.55 < \gamma < -0.25$ & 1  & -0.4 & 0.01 & 5 & 0.938257 & 54 & 0.973615 & 0.00274\\

$-0.58 < \gamma < -0.4$ & 1  & -0.48 & -0.05 & 6 & 1.10431 & 58 & 0.960944 & 0.00285\\

$-0.05 < \gamma < 0.5$ & 2  & 0.25 & 0.01 & 3.5 & 0.65924 & 52 & 0.97281 & 0.00288\\

$-0.1 < \gamma < 0.35$ & 2  & -0.09 & -0.05 & 4 & 0.785847 & 50 & 0.955188 & 0.00366\\

\hline

When $\gamma = 0$, & $\alpha \ne 0$, & $p=2$ & & & & & \\
Range of $\alpha$ & $\lambda$ &  $\alpha $ & $\beta$ & $\phi $ &  $\phi_f$ & N & $n_s$ & r \\  
$-0.03 < \alpha < 0.2$ & 1 & 0.05 & 0.05 & 4 & 0.794314 & 41 & 0.974207 & 0.00415\\

$0.2 < \alpha < 1.0$ & 1 & 0.8 & -0.001 &  5 & 0.974634 & 47 & 0.966983 & 0.00343\\

$0.25 < \alpha < 0.7$ & 2 & 0.48 & 0.05 & 3.8 & 0.728853 & 43 & 0.972846 & 0.00246\\

$-0.43 < \alpha < 0.15$ & 2 & -0.08 & -0.001 & 3.5 & 0.693562 & 44 & 0.964678 & 0.00374\\
\hline

\end{tabular}
\caption{\label{table:2a} For $V = V_0 \bigl(\frac{\lambda \phi^p}{1+ \lambda \phi^p}\bigr)$, the e-fold number $N$ and the spectral index parameters $n_s$ and $r$ calculated for a fixed value of $\phi$ and $\alpha$ are presented. Here $\phi$ is rescaled as $\frac{\phi}{M_{Pl}}$.}
\end{table}
\noindent
We have tabulated our results in Table (\ref{table:2a}) for different choices of the potential parameters. With the $RT$ mixing term only, and for $\beta = -0.01(-0.04)$, $\lambda=1(2)$ and setting $\alpha=\gamma=0$, we find the e-fold parameter $N=54(59)$ , the spectral index $n_s = 0.969229(0.964436)$ and the tensor-to-scalar ratio $r=0.00392(0.0029)$, while by taking $\alpha=0$ and $\gamma = -0.3338(0.1087)$ and  $\beta = -0.010(0.001)$, we find $N=47(46)$ and $n_s = 0.96497(0.97333)$ and $r=0.05178(0.024805)$. We observe that the value of $n_s$ and $r$ for fixed $\alpha$, $\gamma$, and a range of $\beta$ ($-0.04<\beta<0.0171$) matches well with the data provided by WMAP and Planck 2018. By including $T$ term along with $RT$ mixing term, we have seen that  $\gamma$ lies between $[-0.55,-0.25], [0.05,0.5]$  for $\beta>0$ and $[-0.58,-0.4], [-0.1,0.35]$ for $\beta<0$ corresponding to $\lambda=1,2$. The range of $\alpha$ we have got by taking $\gamma=0$ as $[-0.03,0.2], [0.25,0.7]$ for $\beta>0$ and $[0.2,1.0],[-0.43,0.15]$ for $\beta<0$ corresponding to $\lambda=1,2$.\\
\begin{figure}[htb]
 \centerline {\psfig{file=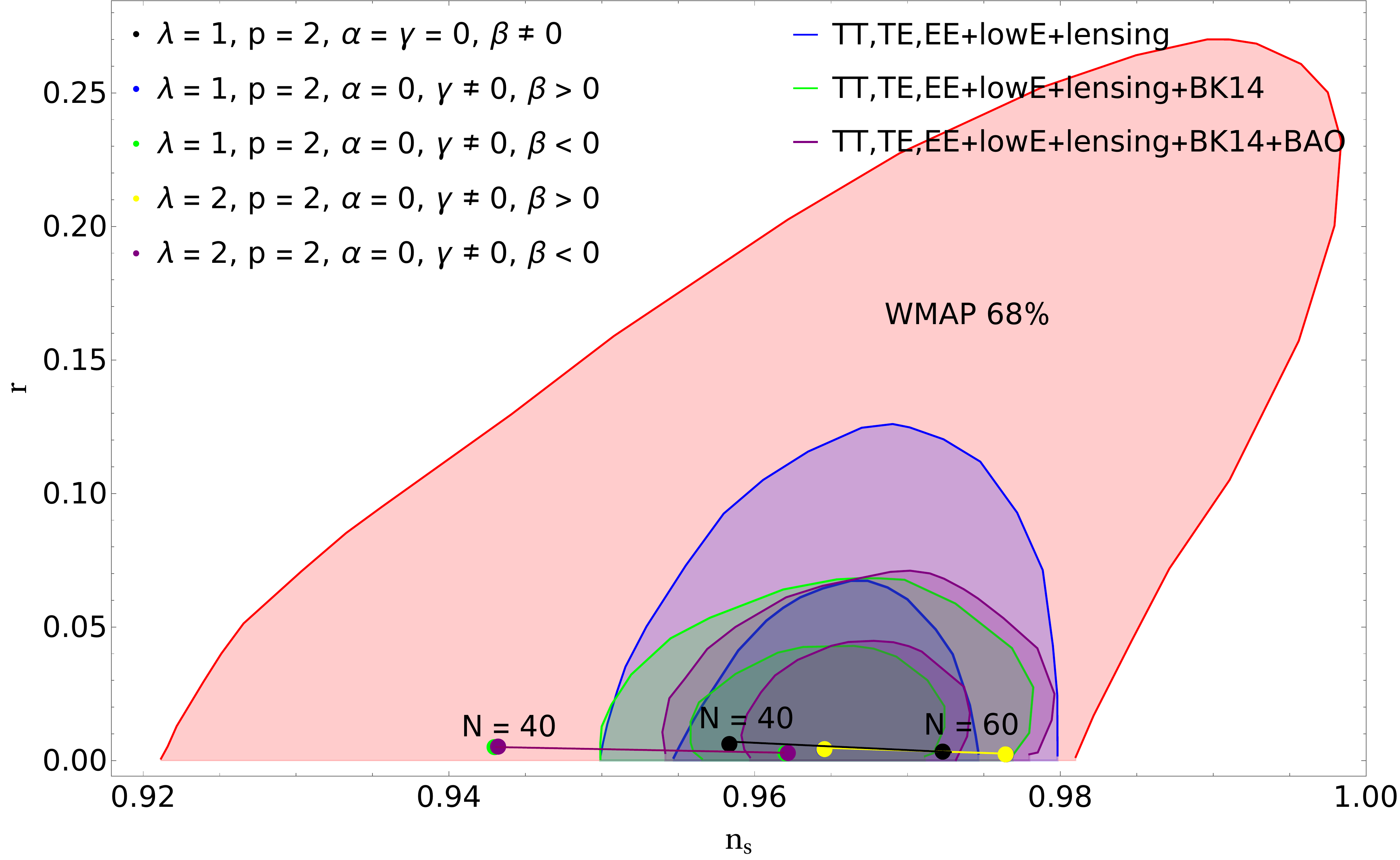,width=8.5cm}\psfig{file=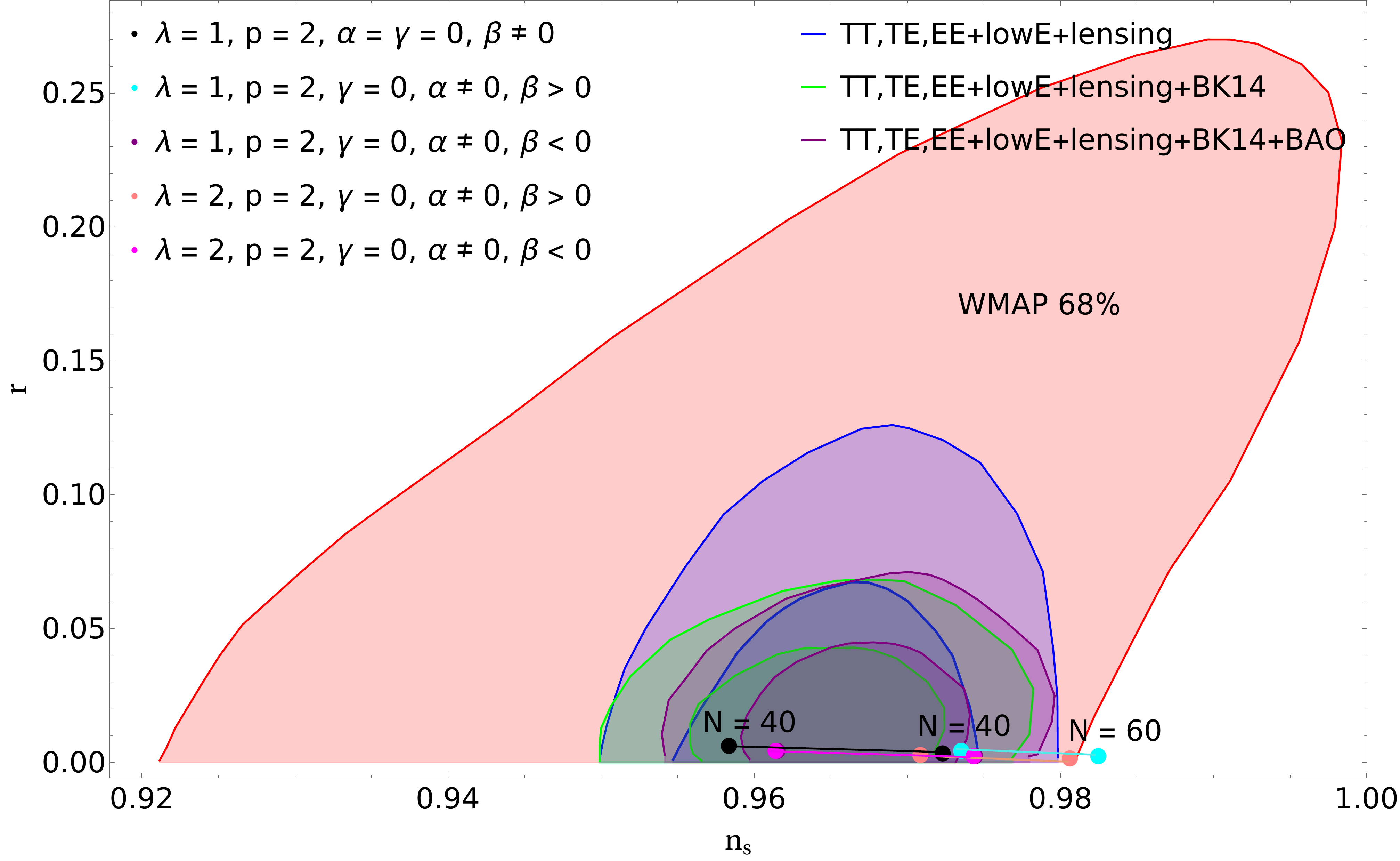,width=8.5cm}}
 \vspace*{3pt}
  \caption{(Color online) Constraints on $n_s$ and $r$ from CMB measurements of different potential. Shaded regions are allowed by WMAP measuremnts, PLANCK alone, PLANCK+BK15, PLANCK+BK15+BAO upto $68\%$ Confidence Level. \label{Plot2}}
\end{figure}
\noindent
Figure \ref{Plot2} represents the contour plot of $n_s$--$r$  for different values of $\alpha$, $\beta$, $\gamma$ and the potential parameters ($\lambda$ and p) for the fractional potential $V= V_0\frac{\lambda\phi^p}{1+\lambda\phi^p}$. Red contour represents the WMAP $n_s - r$ data with $68\%$ C.L. while the blue, purple and green regions represent the Planck 2018 data. Next, we have taken two different values of the potential parameter $\lambda =1$, $2$, and for fixed $p=2$. Then, considered different cases for fixed $\beta$ value (for both cases $\beta>0$ and $\beta<0$) for two cases: (i) $\alpha = 0$ and $\gamma \ne 0$ (ii) $\alpha \ne 0$ and $\gamma = 0$. In the left-hand side plot of Figure \ref{Plot2}, first, we have shown how the $n_s~-~r$ values deviate for non-zero $\beta$ value as we fix $\alpha=\gamma=0$. The black line indicates the plot corresponding to the non-zero $\beta$ value for $N= 40$ and $60$. The yellow and blue lines are overlapped for $\beta>0$ case where we have taken the parameter values as $\lambda =1$, $p=2$, $\alpha =0$, $\gamma \ne 0$ and $\lambda =2$, $p=2$, $\alpha =0$, $\gamma \ne 0$, respectively with $N=40$ and $60$. The identical overlapping scenario for $N=40$ and $60$ can be seen for the $\beta<0$ case where we have taken the parameter values as $\lambda =1$, $p=2$, $\alpha = 0$, $\gamma \ne 0$ and $\lambda =2$, $p=2$, $\alpha = 0$, $\gamma \ne 0$, respectively. On the right-hand side of Figure \ref{Plot2}, we have shown the same black line for the non-zero $\beta$ value for both $\alpha=\gamma=0$ where p is fixed and $\lambda=1$. Then we have considered two different cases as before $\alpha \ne 0$, $\gamma = 0$ when $\beta>0$ and $\beta<0$. In this case, unlike the previous case, we can see that for different $\lambda$'s, distinct lines are coming for $N=40$ and $60$. The blue line and purple line are for the case when $\lambda=1$, $p=2$, $\gamma =0$, $\alpha \ne 0$, $\beta>0$ and $\lambda=1$, $p=2$, $\gamma =0$, $\alpha \ne 0$, $\beta<0$, respectively. The orange and pink lines are for the case when $\lambda=2$ for $\beta>0$ and $\beta<0$, respectively. We can see from the plot that as we switch on the parameter $\alpha$ and $\beta<0$, the $n_s$ value shifts towards the $+3\sigma$ while for the negative $\beta$, the $n_s$ value shifts beyond $+3\sigma$ value. It is important to note that the $n_s$ - r curves cross the boundary of the dark blue contour i.e. falls in the WMAP data range.

\section{Conclusion}

In this manuscript, we have investigated, in detail, a family of non-minimally coupled gravity theories slow-roll inflationary models. In the theory, non-minimal curvature-inflaton couplings result from the assumption that the trace of the energy-momentum tensor $T$ seen in the gravitational action corresponds to inflaton. 
In the present work, we have studied modified $f(R,T)$ gravity with   $f(R,T)=R \bigl(1+\alpha+ \kappa^4\beta T \bigr)+\kappa^2\gamma T $. In particularly, we have focused on the impact of the mixing term $RT$ on inflationary cosmology. In the literature, several $f(R,T)$ models are taken into consideration, where the impacts of $R$ and $T$ are viewed individually. The $RT$ mixing term introduces non-minimal derivative couplings, and higher-order derivative terms can be seen in the equation of motion. By assuming that the inflaton rolls so slowly on its potential that we can neglect the higher-order derivative terms, which makes it quite plausible to write it in Einstein's frame. \\
We have considered two different types of slow-roll inflaton potentials, e.g., $V=V_0\left(1+\ln{\phi}\right)$ and $V=V_0\frac{\lambda\phi^p}{1+\lambda\phi^p}$, have derived the modified slow-roll parameters and different cosmological parameters for these potentials. First, we considered the impact of $RT$ mixing term only by taking $\alpha=\gamma=0$, and later on, we investigated each term by including them one by one together with the $RT$ mixing term. We can observe that although the tensor-to-scalar ratio $r$ value for both the potentials agrees with observational data along with e-fold number $N$ lies between $40-60$, the scalar spactral index $n_s$ value matches with the PLANCK 2018 data for logarithmic potential whereas $n_s$ lies in WMAP region for another potential at least for $N=60$. We get $n_s$ beyond $3\sigma$ region of PLANCK 2018 data by turning off all the modified gravity parameters (i.e. $\alpha=\beta=\gamma=0$) in order to have $N=40-70$ which motivates us towards modified gravity. From Table (\ref{table:2a}), we see that the e-fold $50<N<60$ after including $RT$ mixing term along with $T$ term. \\
We can compare our result with Ref.\cite{Ashmita} where the result without $RT$ mixing term is shown for $V=V_0\frac{\lambda\phi^p}{1+\lambda\phi^p}$. Ref. \cite{Gron} have found the e-fold number $N\approx 31$(a slightly lower value) and the spectral index $n_s\approx 0.99$ for Einstein gravity in the case of Logarithmic inflaton potential, whereas our results presented in Table \ref{table:1a} shows that those are consistent with the observational values of $n_s$ and the desired e-fold number $N$. 
This validates our analysis (with the $RT$ mixing term ) on the estimation of several modified gravity and potential parameters in light of the CMBR observational data.

\section*{Acknowledgments}
PS would like to thank Department of Science and Technology, Government of India for INSPIRE fellowship. Ashmita would like to thank BITS Pilani K K Birla Goa campus for the fellowship support.


\begin{thebibliography}{}

\bibitem{Einstein} A.~Einstein, {\it The Berlin Years: Writings, 1914-1917 }, {\bf 6}, 117,1915
\bibitem{Guth} A. H. Guth, {\it Phys. Rev. D},{\bf 23} (1981) 347–356.
\bibitem{Kolb} E. ~Kolb, M. ~S. ~Turner {\it The Early Universe},(CNC Press), 1994.
\bibitem{Liddle} A. Liddle, {\it An Introduction to Modern Cosmology}, Willey Publication, (2003).
\bibitem{Spergel} D. ~N. ~Spergel {\it et al.}, {\it Physical Review Letter}, {\bf 92}, 20, (2004).
\bibitem{smoot} 
F. G. Smoot, 
{\it AIP Conference Proceedings CONF-981098}, AIP {\bf 476}, 1, (1999).
\bibitem{BICEP} 
P. AR. Ade, 
{\it The Astrophysical Journal}, {\bf 792}, 62, (2014).
\bibitem{PLANCK} Y. ~Akrami, {\it Astronomy \& Astrophysics}, {\bf 641}, 61, (2020).
\bibitem{WMAP} G. ~Hinsaw, {\it Astrophysical Journal Supplement Series}, {\bf 208}, 19, (2013).

\bibitem{Linde1} 
A. D. Linde, 
{\it Phys. Lett.} {\bf B108}, 389-393, (1982).

\bibitem{Linde2} 
A. D. Linde, 
{\it Phys. Lett.} {\bf B129},  177-181, (1983).
\bibitem{Baumann} 
D. Baumann, {\it TASI lecture on Inflation}, arXiv:0907.5424v2 [hep-th].

\bibitem{Kinney} 
W. H. Kinney, {\it TASI lecture on Inflation}, arXiv:0902.1529v2 [astro-ph].
\bibitem{Gron} Ø. Grøn, {\it Universe}, {\bf4}, 15, (2018).

\bibitem{sarkar1} P. ~Sarkar, Ashmita, P. ~K. ~Das, arXiv:2205.05532v2.

\bibitem{Nozari} 
K. Nozari and S. D. Sadatian,
{\it Mod. Phys. Lett.}, {\bf A23}, (2008).

\bibitem{cheong} D. Y. Cheong, S. M. Lee, S. C. Park 
{\it JCAP}, {\bf 2022}, 029, (2022).

\bibitem{Jin} Y. Jin and S. Tsujikawa, {\it Classical and Quantum Gravity}, {\bf23}, 353 (2006)
\bibitem{Maeda} K. i. Maeda, {\it Class. Quantum Gravity} {\bf3}, 233, (1986).
 \bibitem{Accetta2}F.S. Accetta, D.J. Zoller, M.S. Turner, {\it Phys. Rev. D} {\bf31}, 3046
(1985).
\bibitem{Faraoni} V. Faraoni, 
{\it Phys. Rev. D} {\bf53} (1996),6813 [astro-ph/9602111].
\bibitem{Faraoni2} V. Faraoni, 
{\it Phys. Rev. D} {\bf62} (2000),023504 [gr-qc/0002091].
\bibitem{Dzhunushaliev}V. Dzhunushaliev, V. Folomeev, B. Kleihaus and J. Kunz, 
{\it Eur. Phys. J. C} {\bf74} (2014), 2743 [1312.0225]
\bibitem{Dzhunushaliev1} V. Dzhunushaliev, V. Folomeev, B. Kleihaus and J. Kunz, 
{\it Eur. Phys. J. C} {\bf75} (2015) 157 [1501.00886].
\bibitem{Lobato} R.V. Lobato, G.A. Carvalho, A.G. Martins and P.H.R.S. Moraes, 
{\it Eur. Phys. J. Plus} {\bf134} (2019) 132 [1803.08630].
\bibitem{Chen} C.-Y. Chen and Y.-H. Kung, 
{\it Phys. Dark Univ.} {\bf35} (2022) 100956 [2108.04853].
\bibitem{Parker1} L. Parker, {\it Phys. Rev. Lett.} {\bf21}, 562 (2014)
\bibitem{Parker2} L. Parker, Phys. Rev. D 3, 2546 (1971).
\bibitem{Harko} T. Harko, F.S.N. Lobo, S. Nojiri and S.D. Odintsov, 
{\it Phys. Rev. D} {\bf84}
(2011) 024020 [1104.2669].
\bibitem{Roshan} M. Roshan and F. Shojai, 
{\it Phys. Rev. D} {\bf94} (2016)
044002 [1607.06049].
\bibitem{Bertolami} O. Bertolami, C.G. Boehmer, T. Harko and F.S.N. Lobo, 
{\it Phys. Rev. D} {\bf75} (2007) 104016 [0704.1733].
\bibitem{Bisabr} Y. Bisabr, 2012 arxiv:1205.0328v2 [gr-qc]
\bibitem{Sahoo} P.H.R.S. Moraes, P.K Sahoo, 
{\it Eur. Phys. J. C} {\bf77}, 480 (2017)
\bibitem{Nojiri1} S. Nojiri and S.D. Odintsov, 
{\it Phys. Lett. B} {\bf599} (2004) 137 [astro-ph/0403622].
\bibitem{Allemandi} G. Allemandi, A. Borowiec, M. Francaviglia and S.D. Odintsov, 
{\it Phys. Rev. D} {\bf72} (2005) 063505
[gr-qc/0504057].
\bibitem{Chen3} Che-Yu Chen, Y.~ Reyimuaji, X. Zhang, {\it Phys of Dark Unv.} {\bf 38} (2022), 101130
\bibitem{chen2} X. Zhang, Che-Yu Chen, Y.~ Reyimuaji, {\it Physical Review D},{\bf 105}, 4, (2022), [arXiv: 2108.07546]
\bibitem{Nisha} N. Godani, {\it International Journal of Geometric Methods in Modern Physics}, {\bf 16}, 02, (2018).
\bibitem{Ashmita} Ashmita, P.Sarkar, P. K. Das, {\it IJMPD}, {\bf  2250120} (2022) 1-14 
\bibitem{Ashmita2} Ashmita, P. Sarkar, P. K. Das, arXiv:2210.07788
\bibitem{Lyth} D. H. Lyth, A. Riotto, 
{\it Phys. Rep.}, {bf 314}, 1999,  1–146. 
\bibitem{Guo} R. Y. Guo,X. Zhang,
{\it Eur. Phys. J. C},{\bf 77} 2017, 882.
\bibitem{Hebecker} A. Hebecker, S. C. Kraus, D. Lüst, D, S. Steinfurt, T. Weigand,
{\it Nucl. Phys. B}, {\bf 854} 2012, 509–551.
\bibitem{Eshagli} M. Eshagli, M. Zarei, N. Riazi, A. Kiasatpour, {J. Cosmol. Astropart. Phys.}, 2015, 037, [arXiv:1505.03556].
\end{thebibliography}
\end{document}